\documentclass[sigconf]{acmart}
\usepackage{soul}

\AtBeginDocument{%
  }

\usepackage{multirow}
\usepackage{cleveref}
\usepackage{makecell}
\usepackage{bm}
\usepackage{balance}

\usepackage{pifont}
\newcommand{\xmark}{\ding{55}}%

\copyrightyear{2025}
\acmYear{2025}
\setcopyright{rightsretained}
\acmConference[RecSys '25]{Proceedings of the Nineteenth ACM Conference on Recommender Systems}{September 22--26, 2025}{Prague, Czech Republic}
\acmBooktitle{Proceedings of the Nineteenth ACM Conference on Recommender Systems (RecSys '25), September 22--26, 2025, Prague, Czech Republic}\acmDOI{10.1145/3705328.3759318}
\acmISBN{979-8-4007-1364-4/2025/09}

\begin{document}

\title{How Fair is Your Diffusion Recommender Model?}


\author{Daniele Malitesta}
\orcid{0000-0003-2228-0333}
\affiliation{%
  \institution{Université Paris-Saclay, CentraleSupélec, Inria}
  \city{Gif-sur-Yvette}
  \country{France}
}
\email{daniele.malitesta@centralesupelec.fr}

\author{Giacomo Medda}
\orcid{0000-0002-1300-1876}
\affiliation{%
  \institution{University of Cagliari}
  \streetaddress{Via Università, 40}
  \city{Cagliari}
  \country{Italy}
  \postcode{09124}
}
\email{giacomo.medda@unica.it}

\author{Erasmo Purificato}
\authornote{The view expressed in this paper is purely that of the author and may not, under any circumstances, be regarded as an official position of the European Commission.}
\orcid{0000-0002-5506-3020}
\affiliation{%
  \institution{European Commission,\\Joint Research Centre (JRC)}
  \city{Ispra}
  \country{Italy}
}
\email{erasmo.purificato@ec.europa.eu}

\author{Mirko Marras}
\orcid{0000-0003-1989-6057}
\affiliation{%
  \institution{University of Cagliari}
  \streetaddress{Via Università, 40}
  \city{Cagliari}
  \country{Italy}
  \postcode{09124}
}
\email{mirko.marras@acm.org}

\author{Fragkiskos D. Malliaros}
\orcid{0000-0002-8770-3969}
\affiliation{%
  \institution{Université Paris-Saclay, CentraleSupélec, Inria}
  \city{Gif-sur-Yvette}
  \country{France}
}
\email{fragkiskos.malliaros@centralesupelec.fr}

\author{Ludovico Boratto}
\orcid{0000-0002-6053-3015}
\affiliation{%
  \institution{University of Cagliari}
  \streetaddress{Via Università, 40}
  \city{Cagliari}
  \country{Italy}
  \postcode{09124}
}
\email{ludovico.boratto@acm.org}

\renewcommand{\shortauthors}{Malitesta et al.}

\newcommand{\etal}{\textit{et al}. }
\newcommand{\ie}{\textit{i}.\textit{e}., }
\newcommand{\eg}{\textit{e}.\textit{g}., }

\begin{abstract}
    Diffusion-based learning has settled as a rising paradigm in generative recommendation, outperforming traditional approaches built upon variational autoencoders and generative adversarial networks. Despite their effectiveness, concerns have been raised that diffusion models - widely adopted in other machine-learning domains - could potentially lead to unfair outcomes, since they are trained to recover data distributions that often encode inherent biases. Motivated by the related literature, and acknowledging the extensive discussion around bias and fairness aspects in recommendation, we propose, to the best of our knowledge, the first empirical study of fairness for \textit{DiffRec}, chronologically the pioneer technique in diffusion-based recommendation. Our empirical study involves \textit{DiffRec} and its variant \textit{L-DiffRec}, tested against nine recommender systems on two benchmarking datasets to assess recommendation utility and fairness from both consumer and provider perspectives. Specifically, we first evaluate the utility and fairness dimensions \textit{separately} and, then, within a multi-criteria setting to investigate whether, and to what extent, these approaches can achieve a \emph{trade-off} between the two. While showing worrying trends in alignment with the more general machine-learning literature on diffusion models, our results also indicate promising directions to address the unfairness issue in future work.
    The source code is available at \url{https://github.com/danielemalitesta/FairDiffRec}.
\end{abstract}


\maketitle

\section{Introduction}
Research in recommender systems (RSs) has recently been shaped by the latest advances in deep learning, \eg graph neural networks~\cite{DBLP:journals/csur/WuSZXC23}, language models~\cite{balloccu2023knowledge,DBLP:journals/www/WuZQWGSQZZLXC24}, and diffusion models~\cite{DBLP:conf/sigir/WangXFL0C23}. While the first two families have already received significant attention in the literature, diffusion-based RSs are still in early stages.

As of today, diffusion models~\cite{DBLP:conf/nips/HoJA20, DBLP:journals/csur/YangZSHXZZCY24} have settled as the leading solution in generative artificial intelligence (AI)~\cite{DBLP:journals/cacm/GoodfellowPMXWO20} and revolutionized the paradigm of generating new (but realistic) data with several applications, such as computer vision~\cite{DBLP:conf/cvpr/RombachBLEO22, DBLP:conf/nips/HoJA20}, natural language processing~\cite{DBLP:conf/nips/AustinJHTB21, DBLP:conf/nips/LiTGLH22}, and graph modeling~\cite{DBLP:conf/aistats/NiuSSZGE20, DBLP:conf/ijcai/LiuFLLLLTL23}, and domains, such as chemistry and bioinformatics~\cite{DBLP:conf/iclr/XuY0SE022, DBLP:conf/icml/HoogeboomSVW22}. The core principle behind diffusion models lies in their twofold learning procedure, involving a \textit{forward} process where samples from a real-data distribution are iteratively corrupted by adding noise, \eg Gaussian noise, and a \textit{reverse} process where the model is trained to reconstruct the original data by denoising the corrupted samples.

As observed by~\citet{DBLP:conf/sigir/WangXFL0C23}, the diffusion model framework is well-suited for recommendation settings, as it can learn to denoise users' implicit feedback containing noisy interactions, \ie false positives.
This approach, called \emph{DiffRec}, shows improved performance with respect to traditional generative RSs, \eg those based on variational autoencoders (VAEs)~\cite{DBLP:conf/www/LiangKHJ18} and generative adversarial networks (GANs)~\cite{DBLP:conf/sigir/WangYZGXWZZ17}, by addressing the limited representational power of the former and the unstable training procedures of the latter.
Similarly, other works proposed to leverage diffusion models for specific recommendation tasks such as point-of-interests~\cite{DBLP:journals/tois/QinWJLZ24}, sequential~\cite{DBLP:journals/tois/LiSL24, DBLP:conf/aaai/MaXMCZLK24}, or knowledge graph-based recommendation~\cite{DBLP:conf/wsdm/JiangYXH24}.

Diffusion models have achieved widespread success in many applications and domains; however, as a relatively recent development in machine learning (ML), they continue to face several open challenges. For instance, some works have raised concerns regarding the \emph{unfair} outcomes diffusion models could lead to. As observed in generative AI, \emph{bias} is intrinsic in real-world data, and diffusion models might amplify this signal while learning the underlying \emph{biased} data distribution.
In their seminal paper, \citet{DBLP:journals/corr/abs-2302-10893} discuss how stable diffusion methods are prone to generate synthetic images that lack diversity, \eg firefighters are only associated with white males. Other works, such as~\citet{DBLP:conf/fat/0001KDLCNHJ0C23}, discuss how textual prompts with trait annotations may produce stereotypical images. Similarly, \citet{DBLP:conf/iccv/0001ZB23} observe bias in the generated images in terms of people color tones and occupations. Interestingly, the same unfairness issues have also been observed in graph diffusion models~\cite{DBLP:conf/iclr/VignacKSWCF23, DBLP:conf/icml/ChenHH023}, where the generation of both the graph topology and the node features~\cite{DBLP:journals/tmlr/LiKP024} further amplifies the natural bias already existing in the graph structure~\cite{DBLP:conf/nips/KoseS24}. Recent papers have proposed possible countermeasures, such as introducing fairness-guided instructions during the data generation process~\cite{DBLP:journals/corr/abs-2302-10893}, obfuscating the sensitive attributes~\cite{DBLP:conf/aaai/Choi0K0P24}, or adopting an ad-hoc fine-tuning~\cite{DBLP:conf/iclr/ShenDPLWK24, DBLP:conf/nips/KoseS24}.

In this light, a question arises: ``\textit{To what extent do diffusion models exacerbate or replicate unfairness patterns within recommender systems?}''
This question is pertinent given the large relevance that \emph{bias} and \emph{fairness} have held in recommendations for decades~\cite{DBLP:conf/um/AnelliBNJP22, DBLP:journals/tois/MansouryAPMB22, DBLP:conf/kdd/SunGZZRHGTYHC20, DBLP:conf/sigir/0002WLCZDWSLW22,DBLP:journals/ipm/BorattoFMM23}. Thus, we consider it imperative to evaluate the performance of popular diffusion-based RSs rigorously,
with a specific focus on potential \emph{unfairness issues}, as there is currently \emph{no literature on this topic}, despite a well-documented history of bias and unfairness affecting other ML-based RSs.
As diffusion-based recommender models are still in their pioneer era, a thorough investigation of their performance can help identify and address observed pitfalls.

To this end, in this work, we select one of the pioneer approaches in diffusion-based recommendation, \emph{DiffRec}~\cite{DBLP:conf/sigir/WangXFL0C23} (and its variant \emph{L-DiffRec}), to answer the following research question: ``\textit{Is DiffRec a fair recommender system compared to the state-of-the-art?}'' To conduct our investigation, we consider two datasets (with users' attribute data and binary gender label as the sensitive attribute) and nine additional state-of-the-art recommenders.
Our purpose is to benchmark their performance on six recommendation metrics accounting for \emph{utility}, \emph{consumer fairness}~\cite{DBLP:conf/ecir/BorattoFMM22,DBLP:conf/ecir/AnelliDNMPP23,DBLP:journals/air/VassoyL24}, and \emph{provider fairness}~\cite{DBLP:conf/ecir/BorattoFFMM24,DBLP:conf/sigir/NaghiaeiRD22,DBLP:journals/tois/MansouryAPMB22}.
Concretely, we run a twofold analysis, where we first monitor each dimension \emph{separately}, and then in a \emph{multi-criteria} setting. This twofold perspective can help identify patterns by considering all actors in the recommendation scenario~\cite{DBLP:conf/ecir/AnelliDNMPP23}.

Our results depict a worrying scenario, confirming that \emph{unfairness concerns already observed for other ML tasks do indeed extend to the recommendation task}, when leveraging diffusion-based approaches.
Nonetheless, introducing additional components in the diffusion-based process, as for \emph{L-DiffRec}, seems to mitigate the observed unfairness; indeed, these findings pave the way to novel future directions where fairness is preserved in diffusion-based recommendation. Despite its \emph{preliminary nature}, to the best of our knowledge, this is the first attempt to provide a rigorous, fairness-aware evaluation of diffusion-based recommendation methods.

\section{Background}
This section provides the necessary notions regarding diffusion-based recommendation (with a special focus on \emph{DiffRec}~\cite{DBLP:conf/sigir/WangXFL0C23}), and consumer fairness and provider fairness in recommendation.

\subsection{Diffusion Recommender Model}
Building upon diffusion models, \emph{DiffRec}~\cite{DBLP:conf/sigir/WangXFL0C23} proposes to treat the recommendation task as a denoising process of historical user-item interactions, which may be overly noisy (\ie contain false positives).
In detail, the model is trained to first corrupt users' interactions through noise (\textit{forward}) and then reconstruct them via an ad-hoc denoising procedure (\textit{reverse}).
Unlike traditional diffusion models for images, users' historical interactions are not directly corrupted into pure noise, but the ratio of added noise decreases consistently.

Let $\mathcal{U}$ and $\mathcal{I}$ be the sets of users and items in the system, respectively, with $\mathbf{x}_u \in \{0, 1\}^{|\mathcal{I}|}$ as the vector of user-item interactions for user $u$. Each vector entry $\mathbf{x}_{u,i} = 1$ denotes user $u$ interacted with item $i$, otherwise $\mathbf{x}_{u,i} = 0$. Then, the forward process, in a Markov chain process, corrupts the user interaction vector into $T$ steps by gradually adding Gaussian noise. The formulation for any step $0 \leq t \leq T$ is: $q(\mathbf{x}_t | \mathbf{x}_{t- 1}) = \mathcal{N}(\mathbf{x}_t; \sqrt{1 - \beta_t}\mathbf{x}_{t-1}, \beta_t\bm{I})$, where (with a notation abuse) we indicate $\mathbf{x}_0 = \mathbf{x}_u$, $\mathbf{x}_0 \in \mathbb{R}$, at the initial step, $\mathcal{N}$ stands for the Gaussian distribution, $\beta_t \in (0, 1)$ controls the magnitude of the Gaussian noise added at each step $t$, and $\bm{I}$ is the standard deviation of the distribution. Conversely, the reverse process involves training any ML model, \eg a neural network, to approximate the mean and standard deviation of the data distribution and recover the original user interaction vector. Again, for any step $t$, we have: $p_\theta(\mathbf{x}_{t - 1}|\mathbf{x}_t) = \mathcal{N}(\mathbf{x}_{t-1}; \bm{\mu}_\theta(\mathbf{x}_t, t), \bm{\Sigma}_\theta(\mathbf{x}_t, t))$, where $\bm{\mu}_\theta(\mathbf{x}_t, t)$ and $\bm{\Sigma}_\theta(\mathbf{x}_t, t)$ are the learned mean and standard deviation through the ML model parametrized by $\theta$.

Furthermore, to address large-scale recommendation (a pretty common scenario in real-world applications), \citet{DBLP:conf/sigir/WangXFL0C23} also propose \emph{L-DiffRec}, a lighter-weight version of \emph{DiffRec}. This variant first clusters pre-trained item embeddings, \eg from LightGCN, into $C$ clusters. Afterward, the initial user interaction vector $\mathbf{x}_0$ is split into $C$ sub-vectors $\{\mathbf{x}_0^{c}\}_{c=1}^C$ according to the corresponding $c$-th item cluster. Then, a VAE is trained to compress each user interaction sub-vector $\mathbf{x}_0^c$, so that the diffusion, \ie forward/reverse processes, occurs in a low-dimensional latent space, ideally in parallel computation across all splits. Finally, the original user interaction vector is reconstructed through the decoding process of the previous VAE.

Note that in~\citet{DBLP:conf/sigir/WangXFL0C23}, the authors eventually propose a second variant to \emph{DiffRec}, called \emph{T-DiffRec}, which can encode the temporal information during the training. However, \emph{T-DiffRec} is specialized for sequential recommendation, which would require a dedicated study to compare its performance with other analogous models.
To this end, we exclude \emph{T-DiffRec} from our experiments, leaving its for future extensions of our study.

Our analysis aims to offer a \emph{preliminary} outline on the fairness performance of diffusion-based RSs.
Although other works have recently contributed with novel diffusion-based solutions~\cite{DBLP:journals/tois/QinWJLZ24, DBLP:journals/tois/LiSL24, DBLP:conf/aaai/MaXMCZLK24, DBLP:conf/wsdm/JiangYXH24}, \emph{DiffRec} could be rightly considered as the \emph{pioneer} approach in this domain for chronological reasons. In light of this, we plan to extend our analysis to other diffusion models in future works.

\subsection{Fairness Evaluation in Recommendation}
Recent studies, both in RS~\cite{DBLP:conf/ecir/BorattoFMM22,DBLP:conf/sigir/NaghiaeiRD22,DBLP:journals/air/VassoyL24,boratto2023counterfactual,medda2024gnnuers} and relatedly user modeling~\cite{DBLP:conf/um/PurificatoMBL25,DBLP:conf/cikm/PurificatoBL22,DBLP:journals/mima/PurificatoBL24}, focused on assessing algorithmic fairness under the notion of \emph{demographic parity}~\cite{DBLP:conf/innovations/DworkHPRZ12,DBLP:conf/kdd/FeldmanFMSV15}.
From a \emph{consumer} perspective~\cite{DBLP:conf/recsys/BorattoFFMM24,DBLP:conf/sigir/NaghiaeiRD22}, we operationalize this notion as the absolute difference in recommendation utility between users' groups.
To this end, we select the respective fairness counterparts of nDCG and Recall, namely $\Delta$nDCG and $\Delta$Recall.
On the other hand, \textit{provider} fairness is estimated as the absolute difference in exposure (\ie how evenly items or groups of items are exposed to users or groups of users) between items' groups ($\Delta$Exp)~\cite{DBLP:conf/sigir/NaghiaeiRD22,DBLP:conf/ecir/BorattoFFMM24} and as the average percentage of long-tail items in the top-$k$ lists (APLT)~\cite{DBLP:conf/ecir/AnelliDNMPP23,DBLP:journals/tois/MansouryAPMB22}. For all metrics except APLT, values closer to zero indicate fairer recommendations. Due to space constraints, we do not delve into the detailed formulations. Interested readers are referred to the cited works for a comprehensive presentation.

\section{Experimental Study}
In this section, we present our experimental study aimed at assessing recommendation utility and fairness in diffusion-based methods. First, we describe the baselines and datasets adopted for the analysis. Then, we provide reproducibility details for the experiments.

\subsection{Baselines} We select nine state-of-the-art recommendation models along with \emph{DiffRec} and \emph{L-DiffRec}: BPRMF~\cite{DBLP:conf/uai/RendleFGS09}, ItemkNN~\cite{DBLP:conf/www/SarwarKKR01}, NeuMF~\cite{DBLP:conf/www/HeLZNHC17}, LightGCN~\cite{DBLP:conf/sigir/0001DWLZ020}, UltraGCN~\cite{DBLP:conf/cikm/MaoZXLWH21}, XSimGCL~\cite{DBLP:journals/tkde/YuXCCHY24}, EASE~\cite{DBLP:conf/www/Steck19}, MultiVAE~\cite{DBLP:conf/www/LiangKHJ18}, and RecVAE~\cite{DBLP:conf/wsdm/ShenbinATMN20}. As reported in~\Cref{tab:baselines}, our selection was guided by the following observations: (i) the baselines are well-established solutions, accepted at top-tier conferences and journals; (ii) they cover a diverse set of recommendation families, such as graph neural networks and generative models; (iii) many of these approaches have been already analyzed from a recommendation fairness perspective.

\subsection{Datasets}
We assess the performance of the selected baselines on MovieLens-1M (\textbf{ML1M})~\cite{DBLP:journals/tiis/HarperK16} and Foursquare Tokyo (\textbf{FTKY})~\cite{DBLP:journals/tist/YangZQ16}, which are frequently adopted in fairness-aware literature in recommendation~\cite{DBLP:journals/air/VassoyL24,DBLP:conf/recsys/BorattoFFMM24,DBLP:journals/kbs/LiuMSWH21}. While ML1M is a popular recommendation dataset that collects movies and users' preferences on them, FTKY includes users' check-ins at points of interest in Tokyo. In both datasets, users' sensitive information comes in a binary gender label.
Thus, fairness analyses on the consumer side adopt the gender attribute included in both datasets (where female is the protected group), whereas provider fairness is assessed on a binary partition of the items' set, namely \emph{short-head} (popular) and \emph{long-tail} (niche) items (where niche items are the protected group). Following a standard procedure in the literature, as in~\cite{DBLP:conf/ecir/BorattoFFMM24,DBLP:conf/sigir/NaghiaeiRD22,DBLP:journals/tois/MansouryAPMB22}, we select the 20\% most popular items as short-head and the rest as long-tail. Dataset statistics are reported in \Cref{tab:datasets}.

\subsection{Reproducibility Details}
Datasets are split into train, validation, and test sets (70\%/10\%/20\%) following the temporal hold-out method adopted in~\cite{DBLP:conf/sigir/WangXFL0C23,DBLP:conf/recsys/BorattoFFMM24}.
We use the RecBole framework~\cite{DBLP:conf/cikm/ZhaoHPYZLZBTSCX22} to train and evaluate all baseline models. For the two diffusion-based recommender systems, we adopt their original code implementations\footnote{\url{https://github.com/YiyanXu/DiffRec}. Accessed \today.}. Baseline models are trained using a grid search over hyperparameters, guided by the settings in the original papers. The configuration yielding the highest Recall@20 on the validation set is selected for each model.

\begin{table}[!t]
    \centering
    \small
    \caption{Methods benchmarked in this study, along with their publication year, venue, and references to prior works that analyzed them from a fairness perspective.}
    \label{tab:baselines}
    \begin{tabular}{lccc}
        \toprule
        \textbf{Models} & \textbf{Year} & \textbf{Venue} & \textbf{Fairness analyzed in} \\
        \midrule
        ItemkNN~\cite{DBLP:conf/www/SarwarKKR01} & 2001 & WWW & \cite{DBLP:conf/um/AnelliBNJP22,DBLP:journals/ipm/BorattoFMM23,DBLP:conf/ecir/BorattoFMM22} \\
        BPRMF~\cite{DBLP:conf/uai/RendleFGS09} & 2009 & UAI & \cite{DBLP:conf/um/AnelliBNJP22, DBLP:conf/ecir/AnelliDNMPP23, DBLP:journals/tois/MansouryAPMB22, DBLP:conf/kdd/SunGZZRHGTYHC20, DBLP:conf/sigir/0002WLCZDWSLW22,DBLP:conf/ecir/BorattoFMM22} \\
        NeuMF~\cite{DBLP:conf/www/HeLZNHC17} & 2017 & WWW & \cite{DBLP:conf/um/AnelliBNJP22,DBLP:journals/ipm/BorattoFMM23, DBLP:journals/tois/MansouryAPMB22, DBLP:conf/kdd/SunGZZRHGTYHC20,DBLP:conf/ecir/BorattoFMM22}\\
        MultiVAE~\cite{DBLP:conf/www/LiangKHJ18} & 2018 & WWW & \cite{DBLP:conf/um/AnelliBNJP22,DBLP:conf/www/TogashiAS24} \\
        EASE~\cite{DBLP:conf/www/Steck19} & 2019 & WWW & \cite{DBLP:conf/um/AnelliBNJP22}\\
        RecVAE~\cite{DBLP:conf/wsdm/ShenbinATMN20} & 2020 & WSDM & \xmark \\
        LightGCN~\cite{DBLP:conf/sigir/0001DWLZ020} & 2020 & SIGIR & \cite{DBLP:conf/ecir/AnelliDNMPP23, DBLP:conf/recsys/BorattoFFMM24, DBLP:conf/sigir/0002WLCZDWSLW22} \\
        UltraGCN~\cite{DBLP:conf/cikm/MaoZXLWH21} & 2021 & CIKM & \cite{DBLP:conf/ecir/AnelliDNMPP23,DBLP:conf/recsys/BorattoFFMM24} \\ 
        DiffRec~\cite{DBLP:conf/sigir/WangXFL0C23} & 2023 & SIGIR & \xmark\\
        L-DiffRec~\cite{DBLP:conf/sigir/WangXFL0C23} & 2023 & SIGIR & \xmark \\
        XSimGCL~\cite{DBLP:journals/tkde/YuXCCHY24} & 2024 & TKDE & \cite{DBLP:conf/recsys/BorattoFFMM24} \\
        \bottomrule
    \end{tabular}
\end{table}

\begin{table}[!t]
    \centering
    \small
    \caption{Recommendation datasets along with their statistics and information about users' groups and attributes.}
    \label{tab:datasets}
    \begin{tabular}{lrr}
        \toprule
        \textbf{Datasets} & \textbf{ML1M}~\cite{DBLP:journals/tiis/HarperK16} & \textbf{FTKY}~\cite{DBLP:journals/tist/YangZQ16} \\ \cmidrule{1-3}
        \textbf{\# Users} & 6,040 & 7,240 \\
        \textbf{\# Items} & 3,650 & 5,779 \\
        \textbf{\# Interactions} & 1,000,130 & 353,769 \\
        \textbf{Sparsity} & 95.46\% & 99.15\%\\
        \textbf{Gender Repr.} & M : 71.7\%; F : 28.3\% & M : 87.9\%; F : 12.1\% \\
        \textbf{Min. Interactions} & 20 & 20 \\
        \bottomrule
    \end{tabular}
\end{table}

\section{Results and Discussion}
In this section, we aim to answer the following research question: ``\textit{Is DiffRec a fair RS compared to state-of-the-art methods?}". To this end, we first investigate the recommendation utility and fairness dimensions \textit{separately}, and then we inspect the \textit{trade-off} among metrics from both dimensions to assess whether \emph{DiffRec} can deliver recommendations that are both fair and effective.

\begin{table*}[!t]
    \centering
    \caption{Results on top-20 recommendation lists for the selected baselines on ML1M and FTKY when measuring recommendation utility (\ie Recall and nDCG), consumer fairness (\ie $\Delta$Recall and $\Delta$nDCG), and provider fairness (\ie APLT and $\Delta$Exp). For each metric, the arrow direction indicates whether higher ($\uparrow$) or lower ($\downarrow$) values stand for better performance. Boldface and underline refer to the best and the second-to-best results. Metric values are expressed as percentages (\%).}
    \label{tab:fairness}
    \resizebox{\textwidth}{!}{%
    \begin{tabular}{lrrrrrrrrrrrrrr}
        \toprule
        \textbf{Metrics} & \textbf{Random} & \textbf{Pop} & \textbf{BPRMF} & \textbf{ItemkNN} & \textbf{NeuMF} & \textbf{LightGCN} & \textbf{UltraGCN} & \textbf{XSimGCL} & \textbf{EASE} & \textbf{MultiVAE} & \textbf{RecVAE} & \textbf{DiffRec} & \textbf{L-DiffRec} \\ \cmidrule{1-14}
        & \multicolumn{13}{c}{\textit{Movielens-1M (ML1M)}} \\ \cmidrule{1-14}
        Recall ($\uparrow$) & \textit{0.61} & \textit{6.88} & 10.67 & 9.87 & 10.44 & 10.16 & \textbf{10.74} & 10.46 & 6.45 & 8.25 & 9.68 & \underline{10.71} & 10.38 \\
        nDCG ($\uparrow$) & \textit{1.14} & \textit{10.42} & 12.94 & 12.22 & 12.67 & \underline{13.09} & 12.82 & 12.80 & 8.62 & 12.04 & 12.86 & \textbf{13.19} & 12.85 \\
        $\Delta$Recall ($\downarrow$) & \textit{0.01} & \textit{0.82} & 0.43 & 1.08 & 0.32 & 0.19 & \textbf{0.06} & \underline{0.14} & 0.70 & 0.48 & 0.53 & 0.86 & 0.66 \\
        $\Delta$nDCG ($\downarrow$) & \textit{0.19} & \textit{2.54} & 1.57 & 1.98 & 1.54 & 1.43 & \underline{1.37} & \textbf{1.34} & 1.66 & 2.43 & 2.07 & 2.03 & 2.25 \\
        APLT ($\uparrow$) & \textit{79.80} & \textit{0.00} & \underline{12.47} & 8.01 & 11.75 & 8.11 & \textbf{13.73} & 10.50 & 9.00 & 2.63 & 5.38 & 6.18 & 10.81 \\
        $\Delta$Exp ($\downarrow$) & \textit{0.12} & \textit{100.00} &  \underline{86.29} & 91.18 & 87.36 & 91.32 & \textbf{84.98} & 88.78 & 90.54 & 97.02 & 94.10 & 93.67 & 87.26 \\ \cmidrule{1-14}
        & \multicolumn{13}{c}{\textit{Fousquare Tokyo (FTKY)}} \\ \cmidrule{1-14}
        Recall ($\uparrow$) & \textit{0.30} & \textit{7.89} & 12.17 & 11.77 & 12.48 & 11.03 & 12.22 & \underline{13.07} & 6.19 & 10.06 & 12.30 & \textbf{13.10} & 12.48 \\
        nDCG ($\uparrow$) & \textit{0.22} & \textit{5.57} & 10.73 & 10.48 & 10.93 & 9.87 & 10.85 & \underline{11.37} & 5.48 & 9.21 & 10.90 & \textbf{11.53} & 10.77 \\
        $\Delta$Recall ($\downarrow$) & \textit{0.03} & \textit{0.95} & 0.94 & 0.69 & \underline{0.59} & 0.77 & 1.05 & 0.98 & \textbf{0.40} & 0.97 & 1.00 & 0.80 & 0.61 \\
        $\Delta$nDCG ($\downarrow$) & \textit{0.01} & \textit{0.85} & 0.22 & 0.19 & \underline{0.09} & 0.47 & 0.29 & 0.23 & 0.10 & 0.59 & 0.29 & 0.22 & \textbf{0.02} \\
        APLT ($\uparrow$) & \textit{80.00} & \textit{0.00} & 9.93 & \textbf{18.27} & 10.20 & 7.42 & 5.36 & 11.06 & 6.32 & 3.72 & 7.27 & 10.75 & \underline{16.97} \\
        $\Delta$Exp ($\downarrow$) & \textit{0.09} & \textit{100.00} &  89.27 & \textbf{78.91} & 89.39 & 91.87 & 94.32 & 88.25 & 93.56 & 95.75 & 92.03 & 88.57 & \underline{80.85} \\
        \bottomrule
    \end{tabular}}
\end{table*}

\subsection{Uni-Dimensional Analysis}
\Cref{tab:fairness} shows the results of the selected baselines (along with reference models such as Random and Pop) on ML1M and FTKY in terms of recommendation utility (\ie Recall and nDCG), consumer fairness (\ie $\Delta$Recall and $\Delta$nDCG), and provider fairness (\ie APLT and $\Delta$Exp). Note that each metric is computed on the top-20 recommendation lists and is expressed as a percentage (\%). 

On the recommendation \emph{utility} dimension, the results are closely aligned with those observed in the related literature, where graph- and diffusion-based RSs generally dominate over the other baselines on both recommendation datasets. \emph{DiffRec} shows slightly superior performance than its variant \emph{L-DiffRec}. However, it should be noted that the latter is more computationally lightweight. This gap between the two \emph{DiffRec} variants might be due to the diffusion process being conducted in the latent space by \emph{L-DiffRec}.
The VAEs employed by \emph{L-DiffRec} compress the users' preferences signals, potentially leading to information loss and non-perfect reconstruction during decoding. Conversely, the \emph{DiffRec} forward and reverse processes fully capture the discrete preference information, without dealing with the potential loss caused by information bottlenecks.

Moving to the experiments on the \textit{fairness} dimension, our results exhibit distinct trends across the datasets.
On ML1M, graph-based approaches such as UltraGCN and XSimGCL can provide both highly effective and fair recommendations, while the diffusion-based models settle among the least fair approaches among the considered methods.
On FTKY, the fairest RSs are instead the generative ones, such as EASE and \emph{L-DiffRec}, and the traditional ones, such as ItemKNN and NeuMF.
The diverse trends in consumer fairness across ML1M and FTKY might derive from their different domains. Specifically, users' interactions in the movie domain are driven by intrinsically individual preferences towards specific movie categories, such as those sharing the same actors or the same genre. In contrast, venue check-ins in FTKY might derive from going out decided as groups, \eg with friends or partners. Biases in movie preferences may be more strongly amplified by recommender systems compared to biases in venue preferences.

The \textit{consumer unfairness} exhibited by \emph{DiffRec} on ML1M shows that diffusion models are prone to amplify biases, as also reported by other studies. However, a common observation across the datasets is that the \emph{L-DiffRec} variant leads to fairer recommendations than \emph{DiffRec} and the other considered generative models, \eg MultiVAE and RecVAE. Particularly, regarding APLT and $\Delta$Exp, \emph{L-DiffRec} exhibits fairer outcomes on the \textit{provider} side. This might be attributed to the intrinsic clustering operation of \emph{L-DiffRec}, which shifts the learning process from following the common popularity bias to capturing more nuanced patterns across items' categories.

\subsection{Multi-Dimensional Analysis}
As the three analyzed properties, \ie recommendation utility, consumer fairness, and provider fairness, represent different and possibly \textit{contrasting} dimensions of the recommendation scenario, we extend our previous analysis by assessing the trade-off among them, as done in related studies~\cite{DBLP:conf/ecir/AnelliDNMPP23,DBLP:journals/ipm/BorattoFMM23}. Indeed, we believe that this additional assessment may complement the findings from the \textit{uni-dimensional} analysis, motivating and explaining from other complementary perspectives the previous performance trends. 

\begin{figure}[!t]
    \centering
    \includegraphics[width=\columnwidth]{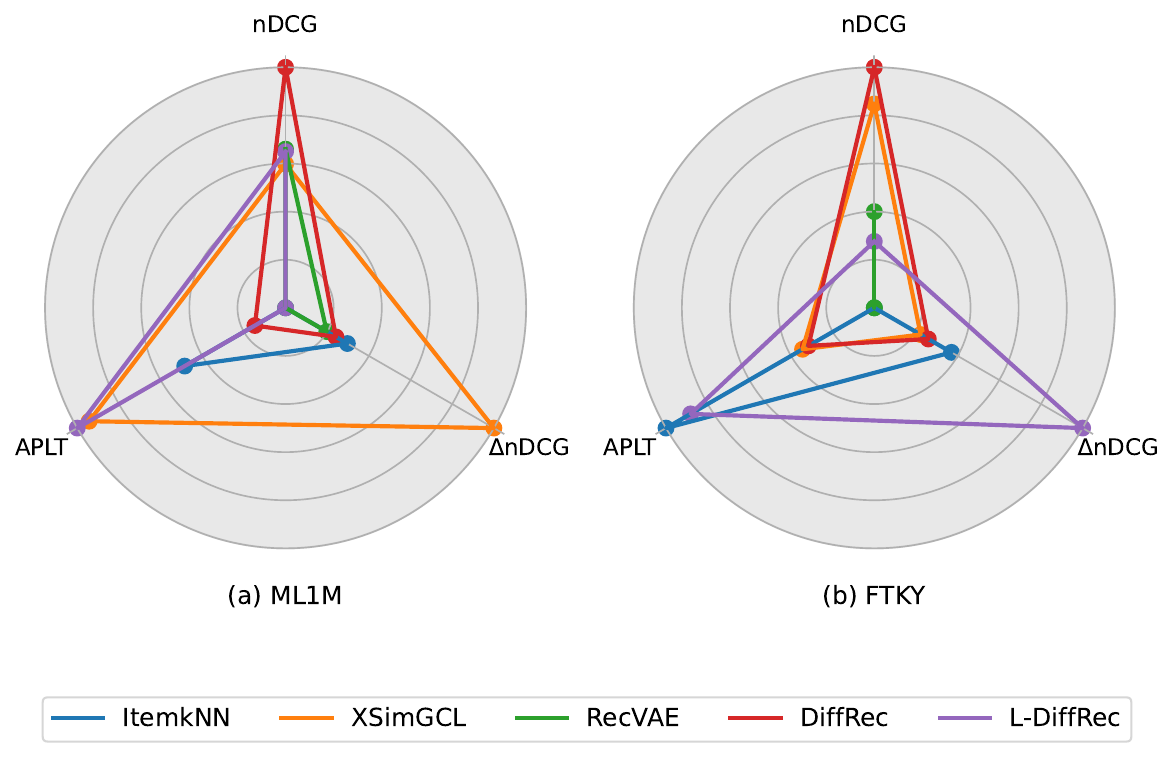}
    \caption{Kiviat diagrams presenting a multi-dimensional analysis of five representative recommendation methods, based on utility (nDCG), consumer fairness ($\Delta$nDCG), and provider fairness (APLT), across the considered datasets. For better visualization, metric values have been normalized in [0, 1], and $\Delta$nDCG has been further rescaled to adhere to the principle ``the higher the better''.}
    \label{fig:radar}
\end{figure}

The Kiviat diagrams in \Cref{fig:radar} represent the performance for both datasets of five representative baselines, \ie ItemkNN, XSimGCL, RecVAE, \emph{DiffRec}, and \emph{L-DiffRec}, on three metrics, \ie nDCG, $\Delta$nDCG, and APLT, accounting for utility, consumer, and provider fairness. To provide an easier interpretation of the plots, we follow \cite{DBLP:conf/ecir/AnelliDNMPP23} and normalize the metrics into [0, 1] through min-max normalization and rescale $\Delta$nDCG to adhere to the principle that ``the higher the values the better the performance''.

On the ML1M dataset, the findings from the \textit{uni-dimensional} analysis are generally confirmed. Indeed, we observe that XSimGCL is the most balanced baseline across all models, being capable of providing an optimal trade-off among all metrics. Conversely, ItemkNN, RecVAE, and DiffRec cannot generally increase one metric side without harming the other ones. Interestingly, though, while ItemkNN tends to maximize fairness over the utility, RecVAE and \emph{DiffRec}, \ie generative and diffusion-based models, do the opposite. This becomes especially evident on \emph{DiffRec}, where the model is largely improving utility at the detriment of provider/consumer fairness. The exception to the trend is \emph{L-DiffRec}, reaching the second-best trade-off soon after XSimGCL. In fact, on FTKY, \emph{L-DiffRec} exhibits even the best trade-off compared to the selected baselines.

In summary, the findings from the \textit{multi-dimensional} analysis confirm the unfairness warnings raised in the \textit{uni-dimensional} analysis but, at the same time, suggest that diffusion-based methods may have room for improvement in achieving fairer recommendations.

\section{Conclusion and Future Work}
This work serves as a preliminary attempt to rigorously assess the performance of diffusion-based RSs, specifically focusing on one of the pioneer approaches, DiffRec, and its variant L-DiffRec. The relevance of our study is underscored by the ML literature, which has recently noted that the intrinsic bias within the data may lead to diffusion models generating unfair outcomes. To this end, we decided to tailor our investigation to the performance of (L-)DiffRec under common fairness properties in RSs.
Specifically, we conducted an extensive analysis involving two recommendation datasets (with sensitive information), nine state-of-the-art RSs, and six recommendation metrics that account for accuracy, consumer, and provider fairness.

Results in two complementary evaluation settings, conducted on single metrics and in a trade-off scenario, empirically confirmed that the same bias amplification caused by diffusion models in ML seems to be advisable also in the recommendation domain (DiffRec). However, when adopting diffusion strategies carefully designed to tailor the recommendation scenario (L-DiffRec), the unfairness issue can be solved and even reverted.
Thus, our empirical study suggests that diffusion models cannot be plugged into RSs as they are, since they would bring (and even exacerbate) the unfairness issues already observed in ML.

However, with careful and tailored modifications to the rationale and models’ architecture (\eg operating in a clustered latent space, as done in L-DiffRec, to reduce noise sensitivity and group-level bias), the issues might be properly addressed.

To confirm our final assumption, future works will aim to extend our study to the other DiffRec variant, T-DiffRec, as well as other diffusion-based RSs~\cite{DBLP:conf/ksem/WalkerZZG022, DBLP:conf/sigir/HouPS24, DBLP:conf/aaai/Choi0K0P24}. While trying to confirm the same observed (un)fairness trends as in this work, we will also devote future investigations to understanding the source of unfairness more carefully, for instance, by examining how the recommendation dataset is modified after the generation process of the diffusion model. Finally, on such theoretical and empirical validations, we plan to exploit the lessons learned from this empirical study and propose tailored strategies to mitigate (when present) the unfairness of diffusion-based recommendation systems.

\begin{acks}
    Daniele Malitesta and Fragkiskos D. Malliaros acknowledge the support of the Innov4-ePiK project managed by the French National Research Agency under the 4th PIA, integrated into France2030 (ANR-23-RHUS-0002).
\end{acks}

%
\balance
\bibliographystyle{ACM-Reference-Format}
\bibliography{references}

\end{document}